\documentclass[aps,prd,superscriptaddress,preprint,tightenlines,nofootinbib]{revtex4}

\usepackage{graphicx} 
\usepackage{dcolumn}  
\usepackage{bm}       
\usepackage{subfigure}  

\begin{document}

\preprint{CLNS 09/2051}       
\preprint{CLEO 09-04}         

\title{\boldmath Search for $\psi(2S) \rightarrow \gamma \eta_{c}(2S)$ via fully reconstructed $\eta_{c}(2S)$ decays}


\author{D.~Cronin-Hennessy}
\author{K.~Y.~Gao}
\author{D.~T.~Gong}
\author{J.~Hietala}
\author{Y.~Kubota}
\author{T.~Klein}
\author{R.~Poling}
\author{P.~Zweber}
\affiliation{University of Minnesota, Minneapolis, Minnesota 55455, USA}
\author{S.~Dobbs}
\author{Z.~Metreveli}
\author{K.~K.~Seth}
\author{B.~J.~Y.~Tan}
\author{A.~Tomaradze}
\affiliation{Northwestern University, Evanston, Illinois 60208, USA}
% ...
\collaboration{CLEO Collaboration}
\noaffiliation


\date{March 5, 2010}

\begin{abstract} 
We report a search for the decay $\psi(2S) \rightarrow \gamma \eta_c(2S)$ in a sample 
of $25.9\times10^6$ $\psi(2S)$ events collected with the CLEO-c detector.  
No signals are observed in any of the 
11 exclusive $\eta_c(2S)$ decay modes studied, or in their sum.  
Product branching fraction upper limits are determined 
as a function of $\Gamma[\eta_c(2S)]$ for the 11 individual modes. 
\end{abstract}

\maketitle


\section{Introduction}

The first radially excited S-wave spin singlet state in the 
charmonium system, $\eta_c(2S)$, 
was observed by the Belle Collaboration in the decay process 
$B^{\pm} \to K^{\pm}~\eta_c(2S), \eta_c(2S) 
\to K^0_S K^{\pm} \pi^{\mp}$ \cite{PhysRevLett.89.102001}. 
It was confirmed by the CLEO \cite{cleo2004fusion} 
and BaBar \cite{babar2004fusion} Collaborations in the two-photon 
fusion process $e^+e^- \to e^+e^-(\gamma \gamma), 
\gamma \gamma \to \eta_c(2S) \to K^0_S K^{\pm} \pi^{\mp}$ and by the 
BaBar Collaboration in the double-charmonium production process 
$e^+e^- \to J/\psi c\bar{c}$ \cite{babar2005jpsicc}.  
These observations, which give an average mass 
$M[\eta_c(2S)] = 3638 \pm 4~{\rm MeV}/c^2$, are inconsistent 
with a previous measurement of $M[\eta_c(2S)] = 3594 \pm 5~{\rm MeV}/c^2$ 
\cite{CB_etac2s} based on an inclusive measurement of the decay process 
$\psi(2S) \to \gamma \eta_c(2S)$.
By measuring the inclusive photon spectrum in a sample of $1.6\times10^6$ 
$\psi(2S)$ decays collected with the CLEO III detector, 
the CLEO Collaboration set an upper limit of 
${\cal{B}}(\psi(2S) \to \gamma \eta_c(2S)) < 0.2\%$ at 90$\%$ 
confidence level (C.L.) for an $\eta_c(2S)$ mass 
of 3594 MeV/$c^2$ \cite{cleo2004photon}.

Theoretical predictions for the branching fraction of 
$\psi(2S) \to \gamma \eta_c(2S)$ based on potential model calculations 
fall in a range of $(0.1-6.2)\times 10^{-4}$ \cite{M1Theory} 
for $M[\eta_c(2S)] = 3638~{\rm MeV}/c^2$. 
A phenomenological prediction, based on assuming that 
the matrix element governing $\psi(2S) \to \gamma \eta_c(2S)$ 
is the same as that for $J/\psi \to \gamma \eta_c(1S)$, is given by 
\begin{equation} 
{\cal B}(\psi(2S) \to \gamma \eta_c(2S)) = \frac{k^3_{\psi(2S)}}{k^3_{J/\psi}}
~\frac{\Gamma_{J/\psi}}{\Gamma_{\psi(2S)}}~{\cal B}(J/\psi \to \gamma \eta_c(1S)),
\label{eq:psi2sToGammaEtac2sPred}
\end{equation}
where $k_{\psi(2S)}$ [$k_{J/\psi}$] is the photon energy for the 
$\psi(2S) \to \gamma \eta_c(2S)$ [$J/\psi \to \gamma \eta_c(1S)$] transition, 
$\Gamma_{\psi(2S)}$ [$\Gamma_{J/\psi}$] is the $\psi(2S)$ [$J/\psi$] full width, 
and ${\cal B}(J/\psi \to \gamma \eta_c(1S)) = (1.72 \pm 0.25)\%$ is the weighted 
average of the value listed by the Particle Data Group (PDG) \cite{PDG2008} 
and a recent CLEO measurement \cite{Mitchell:CLEOjpsitoetac}.  
Using the PDG values for $k_{\psi(2S)}$, $k_{J/\psi}$, $\Gamma_{\psi(2S)}$, and $\Gamma_{J/\psi}$
leads to a prediction of 
${\cal B}(\psi(2S) \to \gamma \eta_c(2S)) = (3.9 \pm 1.1)\times 10^{-4}$.  

In this paper we describe a search for $\eta_c(2S)$ 
production through $\psi(2S) \to \gamma \eta_c(2S)$ using a sample of 
$25.9\times10^6$ $\psi(2S)$ decays collected with the CLEO-c detector.  
We attempt to fully reconstruct the 
$\eta_c(2S)$ in 11 exclusive decay modes:  
$K\bar{K}\pi$ (composed of the decay modes $K^0_S K^{\pm}\pi^{\mp}$ and $K^+K^-\pi^0$), 
$2(\pi^+\pi^-)$, $3(\pi^+\pi^-)$, $K^+K^-\pi^+\pi^-$, 
$K^+K^-\pi^+\pi^-\pi^0$, $K^+K^-2(\pi^+\pi^-)$, 
$K^0_SK^{\pm}\pi^{\mp}\pi^+\pi^-$, 
$\pi^+\pi^-\eta$, $K^+K^-\eta$, $\pi^+\pi^-\eta^{\prime}$, and $\pi^+\pi^-\eta_c(1S)$, 
where the $\eta_c(1S)$ is reconstructed in $K\bar{K}\pi$, $2(\pi^+\pi^-)$, and 
$K^+K^-\pi^+\pi^-$ decays.  
For a specific decay mode $\eta_c(2S) \to X$, the yield of events, $N_{sig}$, 
in a sample of $\psi(2S)$ decays, $N_{\psi(2S)}$, is given by 
\begin{equation} 
N_{sig} = \epsilon N_{\psi(2S)}{\cal B}(\psi(2S) \to \gamma \eta_c(2S)){\cal B}(\eta_c(2S) \to X), 
\label{eq:lighthadronyields}
\end{equation}
where $\epsilon$ is the efficiency for fully reconstructing the $\eta_c(2S)$ hadronic decay 
and the $\sim$50 MeV transition photon.  
The main experimental challenge is background from low-energy photons.  
By searching for exclusive decays to specific final states, it may be possible 
to observe $\eta_c(2S)$ decays through modes other than $K \bar{K} \pi$, 
determine the product branching fractions in Eq. (\ref{eq:lighthadronyields}), 
and obtain improved measurements of $\eta_c(2S)$ properties.

\section{Data Sample and Event Selection}
\label{sect:datasamANDevtsel}

We use an $e^+e^-$ annihilation data sample with an integrated luminosity 
of 51.8 pb$^{-1}$ 
taken at the $\psi(2S)$ mass, $\sqrt{s} = 3.686~{\rm GeV}$.  
The data were produced with the symmetric electron-position beams 
delivered by the Cornell Electron Storage Ring (CESR) 
and collected by the CLEO-c detector \cite{CLEOcDet}.  
CLEO-c is an approximately cylindrically symmetric detector that 
provides a solid angle coverage of 93\%.  
The charged particle tracking system, 
consisting of a six-layer wire vertex detector (ZD) 
and a 47-layer wire drift chamber (DR), 
provides a momentum resolution of 0.6$\%$ for tracks 
with transverse momenta of 1 GeV/$c$.  
An electromagnetic calorimeter (CC) consisting of 7784 cesium iodide crystals 
detects electromagnetic showers with an energy resolution for photons 
of 2.2$\%$ at $E_{\gamma} = 1~{\rm GeV}$ and $\sim5\%$ at 100 MeV.  
Charged particle identification (PID) information is obtained by measuring 
ionization energy loss ($dE/dx$) in the DR and with 
a Ring Imaging Cherenkov (RICH) detector.  The RICH detector is located 
between the DR and CC and covers $|\cos\theta| < 0.83$, 
where $\theta$ is defined with respect to the positron beam.  
All of the detector components described here reside 
within a 1.0 T magnetic field aligned with the beam axis.  

Reconstructed events are required to have the appropriate number 
of charged tracks for the exclusive process being investigated 
and therefore to have zero net charge.  Charged tracks not associated with a 
$K^0_S$ decay are required to have an impact parameter within 5~mm of the 
$e^+e^-$ annihilation interaction point (IP) 
and within 5 cm of the IP along the beam axis.  
The charged pions used to form $K^0_S$ candidates 
are constrained to a common vertex, 
which is required to be displaced from the IP by at least 
3 standard deviations as determined from the net momentum 
of the $\pi^+\pi^-$ pair, and are required to have an invariant mass 
within 10 MeV/$c^2$ of the $K^0_S$ mass.  

PID information from $dE/dx$ and the RICH detector 
is combined to discriminate between pions and kaons.  
Separation is achieved by a requirement on the variable 
$\Delta_{K\pi} = \chi^2_{dE/dx}(\pi) - \chi^2_{dE/dx}(K) 
- 2~{\rm ln}~{\cal L}_{RICH}(\pi) + 2~{\rm ln}~{\cal L}_{RICH}(K)$, 
where $\chi_{dE/dx}(i)$ is the number of standard deviations 
of separation between the measured $dE/dx$ and the mean expectation 
for a particular particle hypothesis, and 
${\cal L}_{RICH}(i)$ is the likelihood for a particular particle hypothesis 
using information from the RICH detector.  
We require kaon candidates to have $\Delta_{K\pi} > 0$, 
while pion candidates are required to have $\Delta_{K\pi} < 0$. 
If there is no information for the track from the RICH detector, 
the number of Cherenkov photons associated with the track is less than 3, 
or the momentum of the track is less than 700 MeV/$c$, 
then only $dE/dx$ information is used.  
If a decay mode includes two charged kaons, 
then only one kaon is required to pass the $\Delta_{K\pi}$ criterion.  

Transition photon candidates are required to be detected in the region 
of the CC that gives the best performance ($|\cos\theta| < 0.81$, ``barrel''), 
to have a lateral shower shape consistent with that of a photon, 
to be not associated with a charged track traversing the CC, and 
to have a minimum energy of 30 MeV.    The $\eta$ candidates are reconstructed 
in the $\eta \to \gamma \gamma$ and $\eta \to \pi^{+}\pi^{-}\pi^{0}$ decay modes, 
with the $\pi^+\pi^-\pi^0$ invariant mass required to be within 10 MeV/$c^2$ 
of the nominal $\eta$ mass [$M(\eta)]$. 
The $\eta^{\prime}$ is reconstructed from the decay process 
$\eta^{\prime} \to \pi^+\pi^-\eta$ with $\eta \to \gamma \gamma$, 
with the $\pi^+\pi^-\eta$ invariant mass required to be within 10 MeV/$c^2$ of $M(\eta^{\prime})$.  
Photon pairs forming a $\pi^0$ ($\eta$) candidate are selected from both the barrel and endcap 
($0.85 < |\cos\theta| < 0.93$) regions of the CC.  
They are required to have a two-photon invariant mass within 
3 standard deviations of the nominal mass, 
approximately $\pm$18 ($\pm$36) MeV/$c^2$ for the $\pi^0$ ($\eta$), 
and are kinematically constrained to the $\pi^0$ ($\eta$) mass 
for subsequent event reconstruction. 

Monte Carlo (MC) simulations, which have been extensively tested with independent data samples, 
are used to determine detector efficiencies and to study backgrounds.
The MC samples are generated with EVTGEN \cite{EVTGEN} 
and a GEANT-based \cite{GEANT} detector simulation.  
Radiation emitted from charged particles, {\it i.e.}, final state radiation (FSR), 
is simulated with the PHOTOS package \cite{PHOTOS}.  
For signal MC samples, the generated angular distribution 
of the vector to vector-pseudoscalar $\psi(2S) \to \gamma \eta_c(2S)$ 
transition is $1+\cos^2\theta$, 
while the $\eta_c(2S)$ is decayed according to phase space.  
MC samples consisting of $259\times 10^6$ generic $\psi(2S)$ decays 
(10 times the data size) and a ``continuum'' sample of 
$e^+e^- \to q\bar{q}~(q = u,d,s)$ events, 
consisting of an integrated luminosity of 259 pb$^{-1}$ 
(5 times the data size), are used to study possible backgrounds.  
The generic $\psi(2S)$ MC sample is generated 
using the available branching fractions 
for the $\psi(2S)$, $\chi_{cJ}$, $J/\psi$, and $\eta_c(1S)$ decays \cite{PDG2007}, 
with unmeasured decay modes simulated by JETSET \cite{JETSET}.  
The continuum sample is generated using JETSET models and has been validated 
with a data sample consisting of 20.6 pb$^{-1}$ collected at 
$\sqrt{s} = 3.671~{\rm GeV}$.

Transitions from the $\psi(2S)$ resonance to other low-lying charmonium 
states are a potentially large background for $\psi(2S) \to \gamma \eta_c(2S)$.  
In particular, decays to the $J/\psi$ via $\pi^+\pi^-$ and $\eta$ transitions and 
transition photons from $\psi(2S) \to \gamma \chi_{cJ}$ need to be suppressed.  
Selection criteria to suppress these decays were designed and efficiencies and 
background-rejection fractions determined with signal and background MC samples, 
respectively.  In what follows, efficiency loss and background rejection 
are determined from the differences in the yields with and without 
the requirement being investigated.

For the hadronic final states $K^+K^-\pi^+\pi^-$, $K^+K^-\pi^+\pi^-\pi^0$, 
$K^0_SK^{\pm}\pi^{\mp}\pi^+\pi^-$, $3(\pi^+\pi^-)$, and $K^+K^-2(\pi^+\pi^-)$, 
events are rejected if any $\pi^+\pi^-$ 
pair originating from the IP has a recoil mass within 20 MeV/$c^2$ of 
$M(J/\psi)$ or if the invariant mass of the other hadrons is 
within 30 MeV/$c^2$ of $M(J/\psi)$.  
Efficiency losses for $K^+K^-\pi^+\pi^-$ are 0.1\% for both criteria, 
and backgrounds are reduced by 10\% and 1\% for the 
$\pi^+\pi^-$ recoil and invariant mass criteria, respectively.  
The efficiency loss for $K^+K^-\pi^+\pi^-\pi^0$ 
is 0.4\% (0.6\%) for the $\pi^+\pi^-$ recoil (invariant) mass criterion, 
while the background is reduced by 26\% (4\%).
Efficiency losses for $K^0_S K^{\pm}\pi^{\mp}\pi^+\pi^-$ are 1.0\% for both criteria, 
and backgrounds are reduced by 57\% and 34\% for the 
$\pi^+\pi^-$ recoil and invariant mass criteria, respectively.  
Efficiency losses for $3(\pi^+\pi^-)$ and $K^+K^-2(\pi^+\pi^-)$ are 5.5\% 
and backgrounds are reduced by two-thirds for the $\pi^+\pi^-$ recoil mass criterion 
while, for the invariant mass criterion,  
the efficiency loss is 5.6\% [6.3\%] and the background is reduced by 33\% [51\%] 
for $3(\pi^+\pi^-)$ [$K^+K^-2(\pi^+\pi^-)$].  

In order to suppress the much more abundant 
$\psi(2S) \to \pi^+\pi^- J/\psi, J/\psi \to \ell^+ \ell^- (\ell = e,\mu)$ 
decays in the $2(\pi^+\pi^-)$ final state, 
events are rejected if the recoil mass of any $\pi^+\pi^-$ 
pair originating from the IP has a value greater than 
$M(J/\psi) - 30~{\rm MeV}/c^2$ = 3067 MeV/$c^2$.  
The efficiency loss for this criterion is 0.7\%, 
while it reduces the background by 96\%.

To suppress $\psi(2S) \to \eta J/\psi$ decays, 
events are rejected if the $\eta$ recoil mass is 
within 40 MeV/$c^2$ of $M(J/\psi)$ for $\pi^+\pi^-\eta$, $\eta \to \gamma \gamma$ 
and within 20 MeV/$c^2$ of $M(J/\psi)$ for $\pi^+\pi^-\eta$, $\eta \to \pi^+\pi^-\pi^0$.  
Efficiency losses are 0.5\% for both $\eta$ decays, 
while backgrounds are reduced by two-thirds.

For the $K^+K^-\pi^0$ final state, 
events are rejected if the recoil mass determined from the higher 
energy photon used in forming the $\pi^0$ candidate is within 20 MeV/$c^2$ 
of $M(\chi_{c2})$ or $M(\chi_{c1})$, or within 30 MeV/$c^2$ of $M(\chi_{c0})$.  
The efficiency loss is 4\%, while the background is reduced by 38\%.  
The same photon recoil mass requirements are used 
for the lower energy photon in $\eta \to \gamma \gamma$ decays 
for the $\pi^+\pi^-\eta$ and $K^+K^-\eta$ final states.  
The efficiency loss is 24\% (27\%) for $\pi^+\pi^-\eta$ ($K^+K^-\eta$), 
while the background is reduced by 83\% (71\%).

The invariant mass of the hadronic decay, $M_{inv}$, for all modes 
is required to be between $M[\psi(2S)]$ and 100 MeV below it, 
{\it i.e.}, $\Delta M \equiv M[\psi(2S)] - M_{inv}$ with $0 < \Delta M < 100~{\rm MeV}/c^2$.  
Requiring $\Delta M > 0~{\rm MeV}/c^2$ rejects events with a direct $\psi(2S)$ decay 
combined with a low-energy shower, while requiring $\Delta M < 100~{\rm MeV}/c^2$ 
rejects hadronic decays of the $\chi_{c2}$ state.  
The efficiency loss is largest for modes with only two charged 
tracks (6.6\% for $K^+K^-\pi^0$, 2.7\% for $K^+K^-\eta,\eta\to\gamma\gamma$, 
and 2.4\% for $\pi^+\pi^-\eta,\eta\to\gamma\gamma$), 
while it is less than 1.7\% for all other modes.  
The background rejection ranges from 21\% for the $K^+K^-2(\pi^+\pi^-)$ mode 
to 67\% for the $2(\pi^+\pi^-)$ mode.

Kinematic fitting is used to optimize signal detection and reject background.  
The sum of the four-momenta of the reconstructed hadronic decay and the 
transition photon candidate is constrained to the initial $\psi(2S)$ 
four-momentum.  The requirement on the $\chi^2$ per degree of freedom 
($\chi^2$/d.o.f.) for this total event fit is optimized mode by mode 
by evaluating the figure of merit $S^2/(S+B)$.  
The accepted signal ($S$) is determined by processing a sample 
of signal MC events that was generated with an assumed branching fraction of 
${\cal B}(\psi(2S) \to \gamma \eta_c(2S)) = 2.6\times 10^{-4}$ \cite{M1Phenom} 
(smaller than our current phenomenological estimate) 
and the arbitrary assumption that the branching fraction for each 
$\eta_c(2S)$ decay to light hadrons is $1\%$.  
The generic $\psi(2S)$ and continuum background MC samples, 
scaled to our data sample size, are used to compute the corresponding 
background ($B$).  The $\chi^2$/d.o.f. requirements 
derived from this study are listed in Table \ref{table:Cuts}.  
In addition, the reconstructed particles originating from the IP 
are constrained to a common vertex, and the $\chi^2$ per degree of freedom 
of this vertex fit is required to be less than 10 for all modes.  

\begin{table}[htp]
\caption[Cuts]
{\label{table:Cuts}
Mode-dependent full event fit $\chi^2$/d.o.f. selection criteria 
and overall signal efficiencies ($\epsilon$) 
for the $\eta_c(2S)$ decaying into light hadrons.  
Efficiencies include statistical uncertainties and 
constituent decay mode branching fractions \cite{PDG2008}.  
We have assumed $\Gamma[\eta_c(2S)] = 14~{\rm MeV}/c^2$.}  
\begin{center}
\begin{tabular}{lcc}
  \hline
  \hline
  Channel & $\chi^2$/d.o.f. & $\epsilon~(\%)$ \\
  \hline
  \hline 

  $K^0_S K^{\pm}\pi^{\mp}$         
    & $<$3.5 & $14.09\pm0.10$ \\ 

  $K^+K^-\pi^0$         
    & $<$4.0 & $17.55\pm0.14$ \\ 

  $K\bar{K}\pi$         
    & --- & $7.63\pm0.04$ \\ 

  $2(\pi^+\pi^-)$  
    & $<$4.5 & $20.48\pm0.16$ \\ 

  $3(\pi^+\pi^-)$  
    & $<$5.0 & $14.22\pm0.14$ \\ 

  $K^+K^-\pi^+\pi^-$
    & $<$4.0 & $19.50\pm0.15$ \\ 

  $K^+K^-\pi^+\pi^-\pi^0$  
    & $<$2.5 &  $8.68\pm0.11$ \\ 

  $K^+K^-2(\pi^+\pi^-)$  
    & $<$4.0 &  $9.93\pm0.11$ \\ 

  $K^0_SK^{\pm}\pi^{\mp}\pi^+\pi^-$  
    & $<$4.0 &  $7.84\pm0.09$ \\ 

  $\pi^+\pi^-\eta, \eta \to \gamma\gamma$
    & $<$2.0 &  $4.03\pm0.04$ \\ 

  $\pi^+\pi^-\eta, \eta \to \pi^+\pi^-\pi^0$
    & $<$3.0 &  $1.65\pm0.02$ \\ 

  $\pi^+\pi^-\eta$
    & --- &  $5.68\pm0.05$ \\ 

  $K^+K^-\eta, \eta \to \gamma\gamma$  
    & $<$3.5 &  $4.55\pm0.05$ \\ 

  $K^+K^-\eta, \eta \to \pi^+\pi^-\pi^0$  
    & $<$5.0 &  $1.92\pm0.02$ \\ 

  $K^+K^-\eta$  
    & --- &  $6.48\pm0.05$ \\ 

  $\pi^+\pi^-\eta^{\prime}$ 
    & $<$3.0 &  $1.42\pm0.02$ \\ 

  \hline
  \hline
\end{tabular}
\end{center}
\end{table}

Additional selection criteria have been developed for suppression of low-energy 
shower backgrounds.  These showers are associated with bremsstrahlung radiation 
emitted from charged pions in the reconstructed hadronic decays (FSR) 
and showers created from nuclear reactions of charged pions and kaons 
in the CC (``split-off'' showers).  The $\chi^2$/d.o.f. requirement 
for the total event fit suppresses some split-off showers, 
but it does not provide effective suppression of FSR since the 
energy momentum is balanced in a fully reconstructed hadronic decay with FSR.  
FSR can be suppressed by requiring that the opening angle between 
a charged pion at the IP and the transition photon candidate 
be greater than some value.  Split-off showers can be suppressed by requiring 
the transition photon candidate to be some distance away from a 
charged track entering the CC.  
The specific selection criteria are optimized using the same $S^2/(S+B)$ 
procedure described above.  

While all decay modes are evaluated 
for additional background shower suppression, 
FSR suppression is found to be useful only for the $2(\pi^+ \pi^-)$ and 
$\pi^+ \pi^- \eta, \eta \to \pi^+ \pi^- \pi^0$ hadronic final states, 
for which the angle between a charged pion and candidate photon 
is required to be greater than 0.376 rad.  
The efficiency loss is 13.0\% (11.1\%) for the $2(\pi^+ \pi^-)$ 
($\pi^+ \pi^- \eta, \eta \to \pi^+ \pi^- \pi^0$) mode, while the 
background is reduced by 41\% (48\%).
Split-off suppression is only applied to the 
$K^+ K^- \pi^+ \pi^-$ and $K^+ K^- \pi^0$ hadronic final states, 
for which the distance between the charged track and candidate photon shower 
is required to be greater than 45 and 35 cm, respectively.
The efficiency loss is 7.7\% (2.2\%) for the $K^+ K^- \pi^+ \pi^-$ 
($K^+ K^- \pi^0$ ) mode, while the background is reduced by 32\% (18\%).

\section{Yield Determination Procedure and Cross-Checks}
\label{sec:procedure}

Searches for the $\psi(2S) \to \gamma \eta_c(2S)$ transition are performed 
by studying the measured shower energy of the transition photon candidate 
without adjustment from the total event kinematic fit.  
Signal yields are determined by performing a binned log-likelihood fit of the 
CC shower energy distribution with a Breit-Wigner function convoluted with 
a MC-determined detector resolution function for the signal shape 
and a background shape composed of the events from the background MC samples 
that pass the event selection criteria.  The mean and width of the 
Breit-Wigner function are fixed to $E_{\gamma} =  48~{\rm MeV}$ and 
$\Gamma_{\gamma} = 14~{\rm MeV}$ \cite{PDG2008}.  
Any monochromatic shower energy 
distribution reconstructed in the CC has a low-side tail 
caused by losses sustained in interactions prior to entering the CC 
and from leakage outside the CsI crystals. For that reason, 
the Crystal Ball function \cite{CBfunction} is used to parameterize 
the detector resolution, with parameters determined from the signal MC samples.  

The procedure for determining the $\psi(2S) \to \gamma \eta_c(2S)$ yields has been 
studied and tested with two closely related processes.  
Reconstructed $\psi(2S) \to \gamma \chi_{c2}, \chi_{c2} \to X$ decays are used 
to test the signal fitting procedure.  To assess the reliability of the 
MC samples for determining the shape of the background shower energy distribution, 
we investigate the process 
$\psi(2S) \to \pi^+ \pi^- J/\psi$ with the $J/\psi$ decaying to $2(\pi^+\pi^-)$, 
$K^+K^-\pi^+\pi^-$, $K^+K^-\pi^0$, or $K^0_SK^{\pm}\pi^{\mp}$.  
These modes have mixtures of final state hadrons very similar to our signal channels 
and no additional photons, so the calorimeter response should closely resemble 
the backgrounds in the $\eta_c(2S)$ signal region.  More details on these studies 
are provided in Ref. \cite{KaiyanThesis}.

\begin{figure}[htbp]
\includegraphics*[width=6.0in]{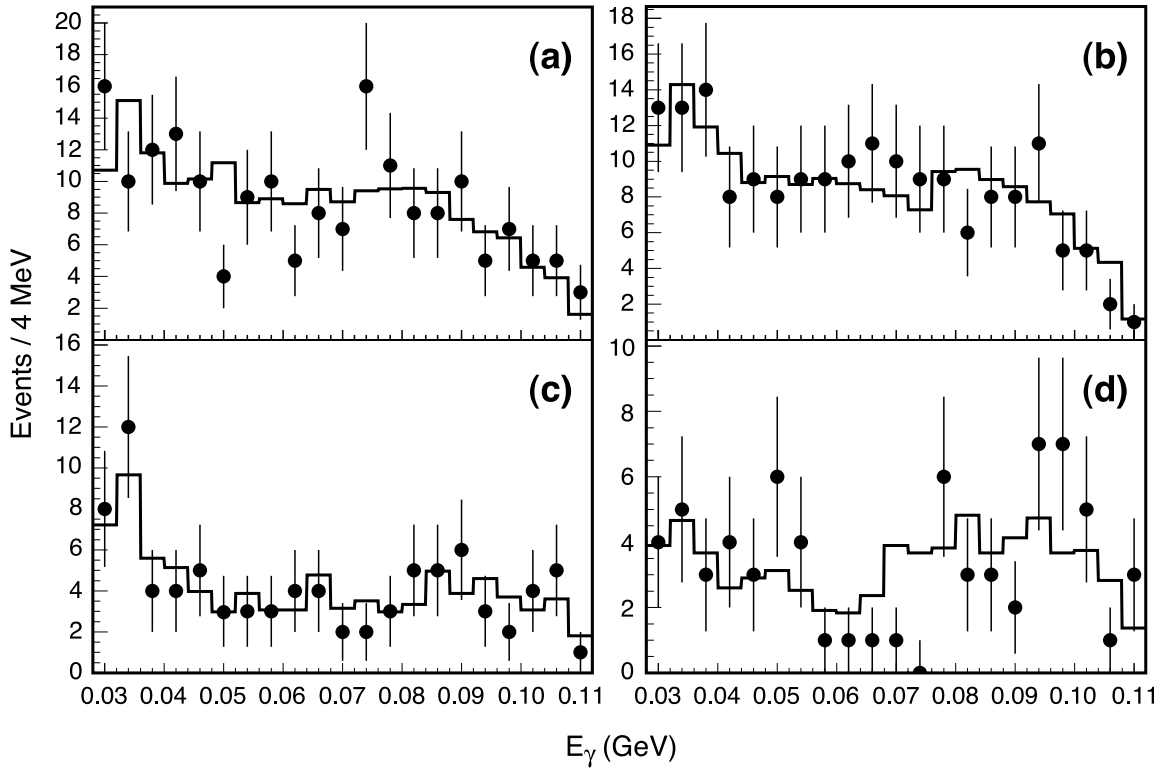}
\caption[Plots from $J/\psi$ bkgd study]
{Distributions of measured shower energy for the decay modes $\psi(2S) \to \pi^+ \pi^- J/\psi$, 
(a) $J/\psi \to 2(\pi^+\pi^-)$, 
(b) $J/\psi \to K^+K^-\pi^+\pi^-$, 
(c) $J/\psi \to K^+K^-\pi^0$, and 
(d)$J/\psi \to K^0_S K^{\pm}\pi^{\mp}$.  
The points are data and the solid histograms are the background MC distributions 
with floating normalization.}  
\label{fig:jpsibkgd}
\end{figure}

For the $J/\psi$ decay study, the previously described event selection criteria 
for the $3(\pi^+ \pi^-)$, $K^+ K^- 2(\pi^+ \pi^-)$, $K^+ K^- \pi^+ \pi^- \pi^0$, 
and $K^0_SK^{\pm}\pi^{\mp} \pi^+ \pi^-$ hadronic final states are applied 
with the exceptions that the $\pi^+ \pi^-$ recoil mass and 
$J/\psi$ hadronic decay suppression criteria are not applied 
and that at least one $\pi^+ \pi^-$ pair is required to have a recoil mass 
within 20 MeV/$c^2$ of $M(J/\psi)$.  
Three different background shapes were studied: a first-order polynomial, 
the energy distribution of showers from the background MC samples that pass 
the $\pi^+ \pi^- J/\psi$ selection criteria (one free parameter for the normalization), 
and events from the same background MC samples partitioned into separate distributions 
for showers identified as being split-off showers and for all others (two free parameters, 
the normalization of each distribution).  
Figure \ref{fig:jpsibkgd} shows the fits of the measured shower energy distributions 
with the backgrounds predicted by the MC with one free parameter 
and no special treatment of split-off showers.  
The $\chi^2$/d.o.f. are 25.1/20, 11.7/20, 11.7/20, and 39.7/20 for the 
$J/\psi \to 2(\pi^+\pi^-), K^+K^-\pi^+\pi^-, K^+K^-\pi^0$, and 
$K^0_S K^{\pm}\pi^{\mp}$ decays, respectively.  
The background MC samples are found to adequately reproduce 
the behavior observed in data, with no clear improvement 
when split-off showers are treated separately.  
Therefore, we use the shower energy distributions from the 
background MC samples with a single normalization parameter 
in fitting the $\eta_c(2S)$ signal region. 

\begin{figure}[htbp]
\includegraphics*[width=3.8in]{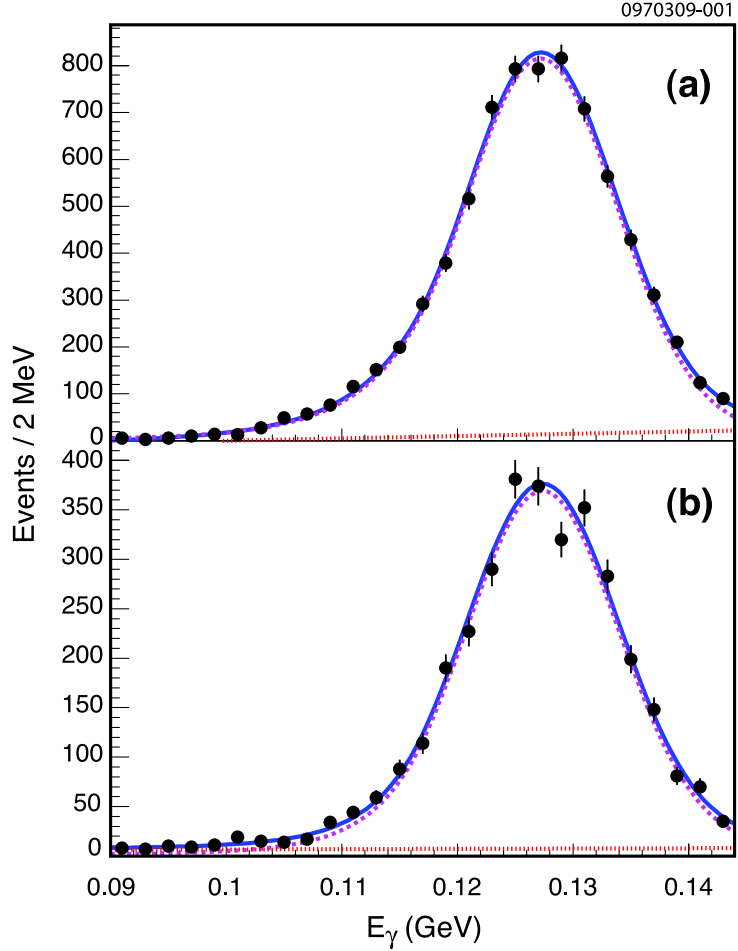}
\caption[Distributions of measured shower energy for $\psi(2S)\to\gamma\chi_{c2}$, 
$\chi_{c2} \to 2(\pi^+\pi^-)$ and  $\chi_{c2} \to K^+K^-\pi^+\pi^-\pi^{0}$ modes.]
{Distributions of measured shower energy for the decay modes $\psi(2S)\to\gamma\chi_{c2}$, 
(a) $\chi_{c2}\to 2(\pi^+\pi^-)$ and (b) $\chi_{c2}\to K^+K^-\pi^+\pi^-\pi^{0}$.  
The points are data; the dashed lines are the signals; 
the dotted lines are the backgrounds; 
and the solid lines are the sums of signal and background.}
\label{fig:chic2examples}
\end{figure}

For the $\psi(2S) \to \gamma \chi_{c2}, \chi_{c2} \to X$ study, 
the $\psi(2S) \to \gamma \eta_c(2S), \eta_c(2S) \to X$ event selection criteria 
are applied, except that the transition photon is required to be 
in the energy range of $90-145~{\rm MeV}$ and no $\Delta M$ requirement is applied.  
Figure \ref{fig:chic2examples} shows the shower energy distributions for the 
$2(\pi^+\pi^-)$ and $K^+K^-\pi^+\pi^-\pi^0$ final states.  
The transition photon signal is well fitted, 
validating the MC-determined resolution function.  
Table \ref{table:Chic2Results} lists the number of signal events 
observed and, for comparison, the number of events expected 
for our $\psi(2S)$ sample using branching fractions from the PDG \cite{PDG2007}.  
The efficiencies are determined from signal MC samples for each hadronic final state, 
where the generated angular distribution of the $\psi(2S) \to \gamma \chi_{c2}$ decay 
is $1+\frac{1}{13}\cos^2\theta$ (which assumes a pure E1 transition \cite{EichtenE1}), 
while the $\chi_{c2}$ is decayed according to phase space.  
The energy resolution for the transition photon, 
determined from the signal MC samples, is $\sim$6.2 MeV.  
Since the substructure of the $\chi_{c2}$ decays and the systematic 
uncertainties of these measurements are not evaluated, 
these yields are presented only as a cross-check of the yield determination procedure 
and not as measurements of the $\chi_{c2}$ decays.

\begin{table}[htbp]
\caption[Chic2Results]
{\label{table:Chic2Results}
Expected and observed yields for the $\psi(2S)\to\gamma\chi_{c2}$ study.  
For each mode, ${\cal B}_{PDG}$ is the value and uncertainty of ${\cal B}(\chi_{c2} \to X)$ 
from the PDG \cite{PDG2007}.  The values in column ``$N_{PDG}$'' are determined from 
$N_{PDG} = {\cal B}(\psi(2S) \to \gamma \chi_{c2}){\cal B}(\chi_{c2} \to X)N_{\psi(2S)}\epsilon$, 
where ${\cal B}(\psi(2S) \to \gamma \chi_{c2}) = (8.1\pm0.4)\%$, 
$N_{\psi(2S)} =  25.9\times 10^{6}$, and $\epsilon$ is the detection efficiency.  
The values in column ``$N_{sig}$'' are fit results and the errors are statistical only.}
\begin{center}
\begin{tabular}{lcccc}
  \hline
  \hline
  Channel & ${\cal B}_{PDG}~(\times 10^{-3})$ & $N_{PDG}$ & $N_{sig}$ & $N_{sig} - N_{PDG}$ \\
  \hline
  \hline
  $K^0_{S}K^{\pm}\pi^{\mp}$ & $0.71 \pm 0.11$ & $262 \pm 43$    & $294 \pm 17$   & $32 \pm 46$ \\
  $K^+K^-\pi^0$           & $0.36 \pm 0.09$ & $192 \pm 49$    & $219 \pm 17$   & $27 \pm 52$ \\
  $2(\pi^+\pi^-)$         & $12.5 \pm 1.6$  & $6947 \pm 953$  & $7215 \pm 119$ & $268 \pm 960$ \\
  $3(\pi^+\pi^-)$         & $8.7 \pm 1.8$   & $3364 \pm 716$  & $6083 \pm 113$ & $2719 \pm 725$ \\
  $K^+K^-\pi^+\pi^-$      & $10.0 \pm 2.6$  & $5226 \pm 1383$ & $4717 \pm 95$  & $-509 \pm 1386$ \\
  $K^+K^-\pi^+\pi^-\pi^0$ & ---             & ---             & $3197 \pm 62$  & ---\\ 
  $K^+K^-2(\pi^+\pi^-)$   & ---             & ---             & $2249 \pm 68$  & ---\\  
  $K^0_{S}K^{\pm}\pi^{\mp}\pi^+\pi^-$ & ---   & ---             & $1453 \pm 54$  & ---\\ 
  $\pi^+\pi^-\eta$          & $0.56 \pm 0.15$  & $109 \pm 29$ & $141 \pm 14$   & $32 \pm 32$ \\
  $K^+K^-\eta$              & $< 0.4$          & $< 89$       & $51.3 \pm 9.1$ & ---\\
  $\pi^+\pi^-\eta^{\prime}$ & $0.59 \pm 0.22$  & $28 \pm 11$  & $3.7 \pm 5.2$  & $-24.3 \pm 12.2$ \\
  \hline
  \hline
\end{tabular}
\end{center}
\end{table}

The numbers of observed events listed in Table \ref{table:Chic2Results} 
are consistent with the PDG \cite{PDG2007}, 
with the exception of the $3(\pi^+\pi^-)$ final state.  
The value listed by the PDG for ${\cal B}(\chi_{c2} \to 3(\pi^+\pi^-))$ comes 
from one measurement \cite{BESIchic2}.  
The same paper presents a measurement for ${\cal B}(\chi_{c2} \to 2(\pi^+\pi^-))$ 
that leads to an expectation of $5112 \pm 1334$ observed events,  
well below both the PDG expectation and our measurement.  
The ratios of the current yields to 
those derived from Ref. \cite{BESIchic2} are $1.4 \pm 0.4$ and $1.8 \pm 0.4$ 
for the $2(\pi^+\pi^-)$ and  $3(\pi^+\pi^-)$ final states, respectively.  
Our observed yield for the $2(\pi^+\pi^-)$ final state 
is consistent with the current PDG value for ${\cal B}(\chi_{c2} \to 2(\pi^+\pi^-))$, 
which is determined from a 28-parameter fit using properties 
of the $\chi_{cJ}$ and $\psi(2S)$.  
While further measurements may clarify the ${\cal B}(\chi_{c2} \to 3(\pi^+\pi^-))$ 
discrepancy, we conclude that the $\psi(2S) \to \gamma \chi_{c2}, \chi_{c2} \to X$ 
study satisfactorily validates our yield determination procedure.

\section{Yield Determinations}

\begin{figure}[htp]
\includegraphics*[width=3.1in]{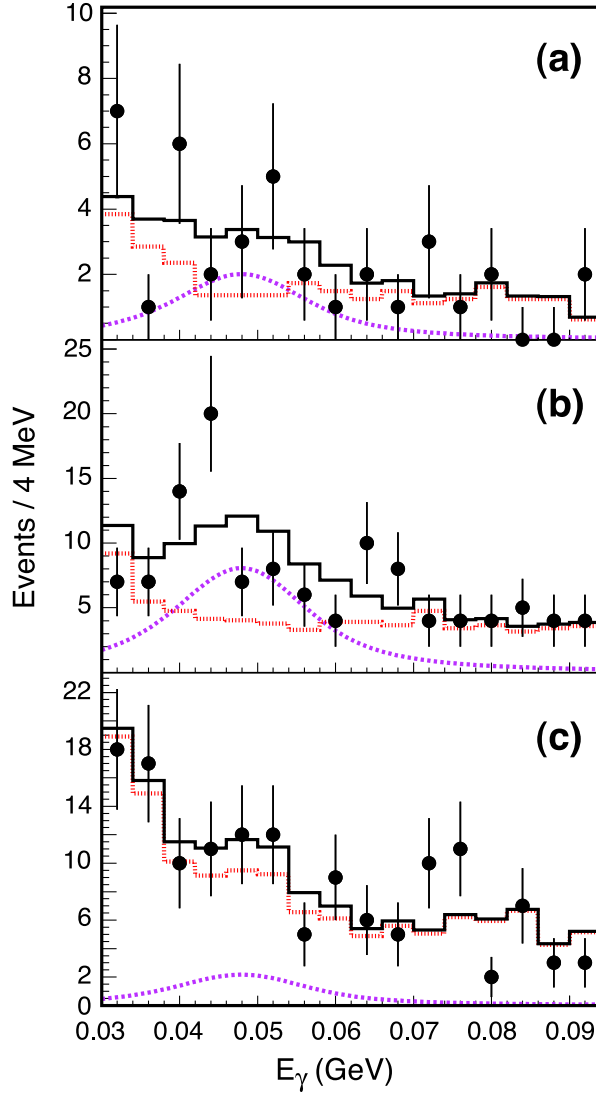}
\caption{Distributions of measured shower energy in the $\eta_c(2S)$ signal region 
for the modes 
(a) $K\bar{K}\pi$, 
(b) $2(\pi^+\pi^-)$, and
(c) $K^+K^-\pi^+\pi^-$.
The points are data; the dashed lines are the signals; 
the dotted histograms are the backgrounds; and 
the solid histograms are the sums of signal and background.}
\label{fig:etac2sPerMode1}
\end{figure}

\begin{figure}[htp]
\includegraphics*[width=6.0in]{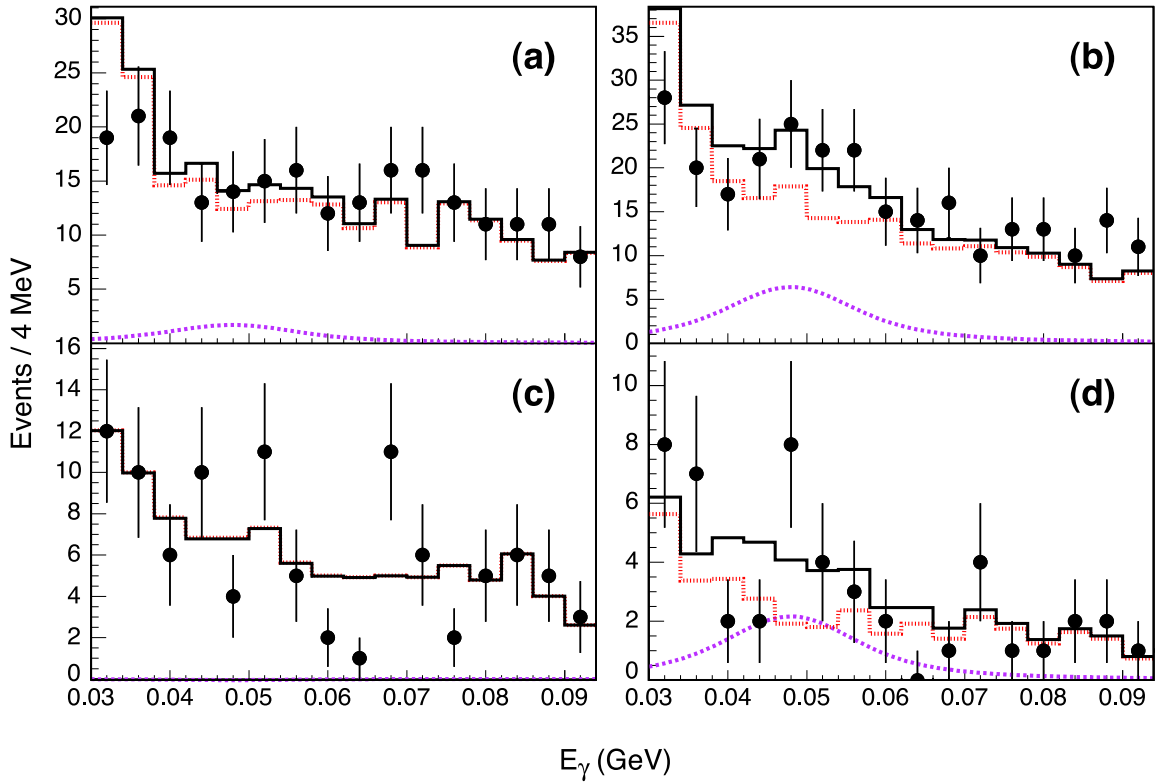}
\caption{Distributions of measured shower energy in the $\eta_c(2S)$ signal region 
for the modes 
(a) $3(\pi^+\pi^-)$, 
(b) $K^+K^-\pi^+\pi^-\pi^0$, 
(c) $K^+K^-2(\pi^+\pi^-)$, and
(d) $K^0_SK^{\pm}\pi^{\mp}\pi^+\pi^-$.
The points are data; the dashed lines are the signals 
[not shown for the $K^+K^-2(\pi^+\pi^-)$ mode since its area is less than zero]; 
the dotted histograms are the backgrounds; and 
the solid histograms are the sums of signal and background.}
\label{fig:etac2sPerMode2}
\end{figure}

\begin{figure}[htp]
\includegraphics*[width=3.1in]{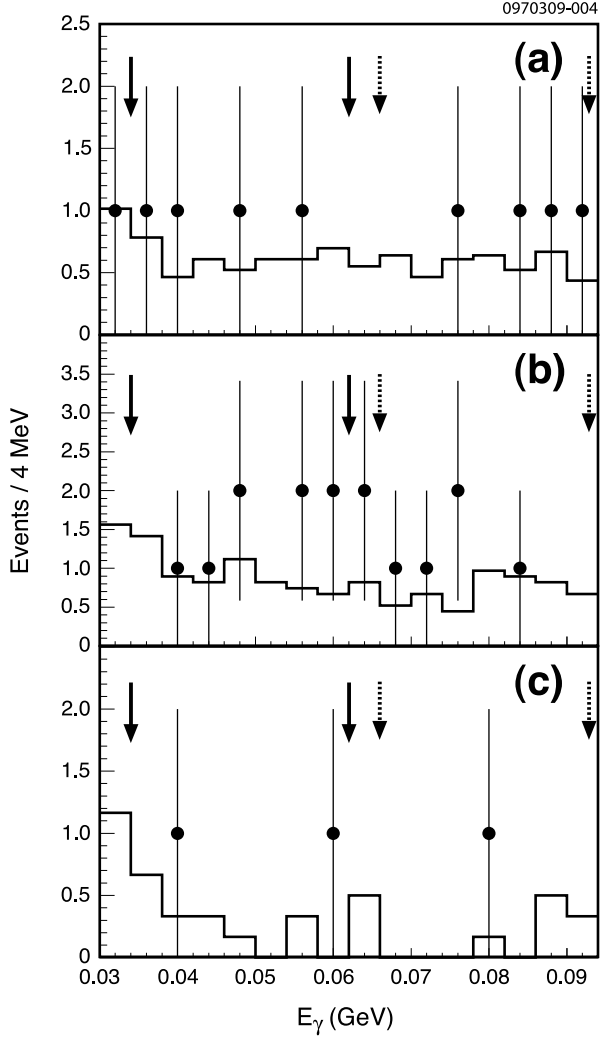}
\caption{Distributions of measured shower energy in the $\eta_c(2S)$ signal region 
for the modes 
(a) $\pi^+\pi^-\eta$, 
(b) $K^+K^-\eta$, and   
(c) $\pi^+\pi^-\eta^{\prime}$. 
The points are data; the solid histograms are the backgrounds; 
the solid arrows enclose the signal region; and 
the dashed arrows enclose the sideband region.}
\label{fig:etac2sPerMode3}
\end{figure}

Figures \ref{fig:etac2sPerMode1} - \ref{fig:etac2sPerMode3} 
show the measured energy distributions of the transition photon candidates 
in the $\eta_c(2S)$ signal region.  
The photon energy resolution in this region, determined from 
the signal MC samples, is $\sim$4.6 MeV.  No significant signal is observed in any mode.  
The $2(\pi^+\pi^-)$ decay mode is the only mode in which an excess 
above background is present.  We have investigated other aspects of the events 
in the signal region and found that this excess, which has a statistical 
significance of slightly more than 3 standard deviations, is most likely 
caused by an upward fluctuation of the background \cite{KaiyanThesis}. 

\begin{figure}[htp]
\includegraphics*[width=4.5in]{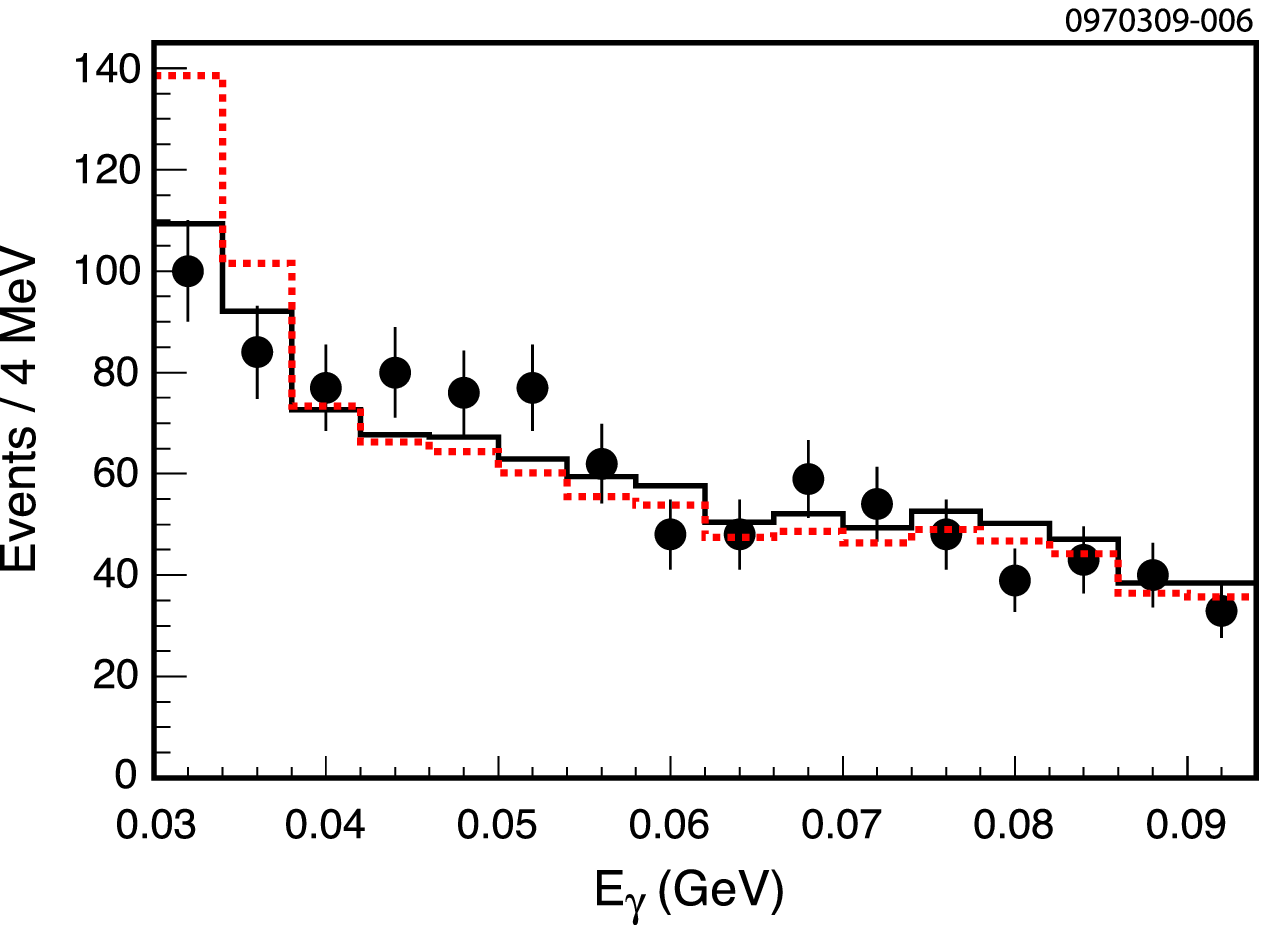}
\caption{Summed distribution of measured shower energy in the $\eta_c(2S)$ signal region 
for all $\eta_c(2S)$ candidates decaying to light hadrons in all ten $\eta_c(2S)$ decay modes.  
The points are data; the dotted histogram is the background without 
special treatment of split-off showers (one free parameter); and the 
solid histogram is the background separately treating the normalizations 
of the split-off and nonsplit-off showers (two free parameters).} 
\label{fig:etac2s10ModeSum}
\end{figure}

For the final states that do not include an $\eta$ decay, 
the signal yield upper limits are determined by finding the value 
corresponding to $90\%$ of the probability distribution determined 
from the measurement, restricted to physically allowed values.  
The yield measurements are listed in Table \ref{table:NonEtaModeResults}.  

\begin{table}[htp]
\caption[NonEtaModeResults]
{\label{table:NonEtaModeResults}
Summary of results and systematic uncertainties for $\eta_c(2S)$ modes 
with yields determined by fitting the transition photon candidate energy distribution.  
The ``$\chi^2$/d.o.f.'' column lists the fit results for 
Figs. \ref{fig:etac2sPerMode1} and \ref{fig:etac2sPerMode2}.  
Central values and 90\% confidence level upper limits are provided 
for the product branching fraction 
$B_{1}B_{2} = {\cal B}(\psi(2S) \to \gamma \eta_c(2S)){\cal B}(\eta_c(2S)\to X)$.  
The columns under ``Systematic uncertainties'' correspond to 
(A) $\Delta M$ criterion, (B) background parameterization, (C) $M[\eta_c(2S)]$ uncertainty, 
(D) signal region, and (E) nonresonant background component, as described in the text.  
Only statistical uncertainties are included in the $N_{sig}$ results, while 
statistical and systematic uncertainties are included in the $B_{1}B_{2}$ results. 
We have assumed $\Gamma[\eta_c(2S)] = 14~{\rm MeV}/c^2$ \cite{PDG2008}.}
\begin{center}
\begin{tabular}{lrccccccccrc}
  \hline
  \hline
          & \multicolumn{1}{c}{$N_{sig}$} & $\chi^2$/d.o.f. & \multicolumn{7}{c}{Systematic uncertainties (\%)} 
          & \multicolumn{2}{c}{$B_{1}B_{2}$ ($\times 10^{-6}$)} \\
  Channel & & & A & B & C & D & E & Other & Total 
          & & (90\% C.L.) \\
  \hline
  \hline 

  $K\bar{K}\pi$         
    & $11.7^{+7.8}_{-7.0}$~~ & 9.7/14
    & ~15.2 & ~16.9 & ~12.2 & ~13.3 & ~8.2 &  5.6 & 30.7 
    & $5.9^{+4.0}_{-3.5}\pm1.8$  & $<$14.5 \\ 

  $2(\pi^+\pi^-)$  
    & $47.9^{+13.6}_{-13.0}$~~ & 14.1/14
    &  ~2.5 &  ~6.1 & ~11.8 & ~11.7 & ~4.4 &  5.0 & 19.1
    & $9.0^{+2.6}_{-2.5}\pm1.7$  & $<$14.6 \\ 

  $3(\pi^+\pi^-)$  
    & $10.1^{+18.1}_{-17.6}$~~ & 11.2/14 
    & ~16.6 & ~20.4 &  ~14.6 & ~12.1 & ~3.6 &  5.1 & 33.0
    & $2.7^{+4.9}_{-4.8}\pm0.9$  & $<$13.2 \\ 

  $K^+K^-\pi^+\pi^-$
    & $12.8^{+15.8}_{-15.6}$~~  & 9.2/14 
    &  ~7.7 & ~32.6 &  ~7.1 & ~13.8 & ~4.4 &  4.5 & 37.2 
    & $2.5^{+3.1}_{-3.1}\pm0.9$  &  $<$9.6 \\ 

  $K^+K^-\pi^+\pi^-\pi^0$  
    & $37.5^{+21.3}_{-20.8}$~~ & 13.8/14 
    & ~22.7 & ~15.2 & ~29.7 & ~24.5 & ~0.9 &  7.2 & 47.8 
    & $16.7^{+9.5}_{-9.3}\pm8.0$  & $<$43.0 \\ 

  $K^+K^-2(\pi^+\pi^-)$~~  
    & $-0.3^{+12.6}_{-12.2}$~~ & 13.2/14
    &  ~0.8 &  ~6.8 & ~14.8 & ~11.0 & ~4.3 &  ~5.3 & 20.8 
    & $-0.1^{+4.9}_{-4.7}\pm0.1$  &  $<$9.7 \\ 

  $K^0_S K^{\pm}\pi^{\mp}\pi^+\pi^-$  
    & $12.9^{+8.3}_{-7.5}$~~  & 11.9/14
    & ~13.2 & ~17.8 & ~16.3 & ~5.6 & ~5.2 &  5.0 & 29.0
    & $6.4^{+4.1}_{-3.7}\pm1.8$  & $<$15.2 \\ 

  \hline
  \hline
\end{tabular}
\end{center}
\end{table}

For the final states that include an $\eta$ decay, 
which have very low statistics, the method 
of Feldman and Cousins \cite{FeldmanCousins} is used 
and only 90\% confidence level upper limits are determined.  
The shower energy distribution is divided into two regions: 
a signal region ($34-62~{\rm MeV}$) corresponding to one full width 
about the $\eta_c(2S)$ mass \cite{PDG2008} 
and a sideband region ($66-94~{\rm MeV}$).  The sideband region is fitted 
with the shape from the background MC samples and the resulting normalization 
is used to compute the number of background events in the signal region, 
listed as $N_{bg}$ in Table \ref{table:EtaModeResults}.  
The number of observed events in the signal region is given as $N_{obs}$.  

\begin{table}[htp]
\caption[EtaModeResults]
{\label{table:EtaModeResults}
Summary of results and systematic uncertainties for $\eta_c(2S)$ modes 
with yields determined by sideband subtraction.  
The product branching fraction is defined as 
$B_{1}B_{2} = {\cal B}(\psi(2S) \to \gamma \eta_c(2S)){\cal B}(\eta_c(2S)\to X)$.  
The columns under ``Systematic uncertainties'' correspond to 
(A) $\Delta M$ criterion, (B) background parameterization, and 
(C) $M[\eta_c(2S)]$ uncertainty, as described in the text.  
Statistical and systematic uncertainties are included in the $B_{1}B_{2}$ results. 
We have assumed $\Gamma[\eta_c(2S)] = 14~{\rm MeV}/c^2$ \cite{PDG2008}.}
\begin{center}
\begin{tabular}{lcccccccc}
  \hline
  \hline
          & $N_{obs}~/~N_{bg}$ & $N_{sig}/\epsilon$ & \multicolumn{5}{c}{Systematic uncertainties (\%)} 
          & ~~$B_{1}B_{2}~(\times 10^{-6})$ \\
  Channel &  & (90\% C.L.) & A & B & C & Other & Total 
          & (90\% C.L.) \\
  \hline
  \hline 

  $\pi^+\pi^-\eta$
    & $4~/~4.3$ &  $<$75.4 
    & ~6.1 & ~46.5 & ~3.0 & 8.1 & 47.7 
    &  $<$4.3 \\ 

  $K^+K^-\eta$  
    & $8~/~6.5$ &  $<$115.7 
    & ~9.2 & ~29.8 & ~3.9 & 7.6 & 32.3 
    &  $<$5.9 \\ 

  $\pi^+\pi^-\eta^{\prime}$ 
    & $2~/~1.8$ & $<$287.9
    & ~8.9 & ~24.5 & ~3.7 & 7.4 & 27.3 
    & $<$14.2 \\ 

  \hline
  \hline
\end{tabular}
\end{center}
\end{table}

Figure \ref{fig:etac2s10ModeSum} shows the summed shower energy distribution 
for the ten $\eta_c(2S)$ decay channels. 
The background distribution in Fig. \ref{fig:etac2s10ModeSum} (dotted histogram) 
has been constructed by adding mode-by-mode background-only fits.  
It shows a visible overestimate compared to the data distribution in the 
lowest energy bins, which we attribute to the modeling of the split-off distribution.  
The alternative treatment using the two-parameter background MC fit described 
in Sec. \ref{sec:procedure} (solid histogram) reproduces the low-energy range 
noticeably better.  
While there may be a small excess in the signal region above the estimated background 
with the two assumed background shapes, the statistical significance is less than 
3 standard deviations and is dependent on the background shape.

\section{Sources of Systematic Uncertainty}
\label{sec:syst}

The sources of systematic uncertainty in our measurements 
of $\eta_c(2S)$ branching fractions 
have been evaluated by reanalyzing the $\psi(2S) \to \gamma \eta_c(2S)$ and 
$\psi(2S) \to \gamma \chi_{c2}$ candidate samples with alternative procedures.  
They are listed in Tables \ref{table:NonEtaModeResults} and \ref{table:EtaModeResults} 
and described in detail below.  
All individual uncertainties not explicitly listed in the tables 
are combined in quadrature and listed as ``Other.''  
The uncertainty in the number of $\psi(2S)$ decays in our sample (2\% \cite{numpsi2s}) 
also affects the branching fraction determinations.  
Other uncertainties in the detection efficiencies that have been evaluated include 
those associated with trigger decisions (1\%), 
reconstruction of the transition photon (2\%) and other particles, and PID.   

The particle reconstruction and PID uncertainties have been estimated using 281 pb$^{-1}$ 
of data collected at the peak of the $\psi(3770)$ resonance \cite{DHadPaper}.  
Reconstruction uncertainties 
are determined by detecting all particles in an event except the 
particle being investigated, determining the efficiency for 
reconstructing the particle in data and MC simulations, and taking the 
difference as the systematic uncertainty.  These studies find uncertainties 
of 0.3\% per charged pion, 0.6\% per charged kaon, 1.8\% per $K^0_S$, 
and 2.0\% for reconstruction of the $\gamma\gamma$ decays of $\pi^0$ and $\eta$.  
The uncertainties associated with PID are determined 
by comparing the efficiency differences between data and MC simulations 
after applying the PID criteria.  The uncertainties are found to be 
0.25\% (0.3\%) per charged pion (kaon).

The uncertainties determined from studying 
$\psi(2S) \to \gamma \chi_{c2}, \chi_{c2} \to X$ decays described below 
are found one by one by removing the selection criterion being investigated, 
redetermining the efficiency-corrected yield, and taking the relative difference 
between this and the nominal case as the systematic uncertainty.
For $J/\psi$ suppression using the $\pi^+ \pi^-$ recoil mass, the uncertainties 
for the $3(\pi^+ \pi^-)$, $K^+ K^- \pi^+ \pi^- \pi^0$, and  $K^+ K^- 2(\pi^+ \pi^-)$ 
final states are 2.4\%, 2.2\%, and 1.3\%, respectively; 
they are less than 1\% for other modes.
For $J/\psi$ suppression using the invariant mass of the decay products,  
the uncertainties for the $3(\pi^+ \pi^-)$ and $K^+ K^- 2(\pi^+ \pi^-)$ 
hadronic final states are 1.0\% and 1.5\%, respectively; 
they are less than 1\% for other modes.
An uncertainty of 2.4\% is assigned to the requirement on the angle 
between the initial pion momentum 
and the candidate photon based on studies of $\chi_{c2} \to 2(\pi^+ \pi^-)$ decays.  
An uncertainty of 0.8\% is assigned to the requirement on the distance between 
a transition photon candidate and the nearest track in the CC based on studies 
of $\chi_{c2} \to K^+ K^- \pi^+ \pi^-$ decays. 
An uncertainty of 2.3\% is assigned for suppressing transition photons 
from $\psi(2S) \to \gamma \chi_{cJ}$ decays in selecting 
$\pi^0$ and $\eta \to \gamma \gamma$ candidates by studying 
$\chi_{c2} \to K^+ K^- \pi^+ \pi^- \pi^0$.
An uncertainty of 1.3\% is conservatively assigned for the requirement on the vertex fit 
by taking the uncertainty from the $\chi_{c2}$ decay mode with the largest discrepancy.
An uncertainty of 4.0\% (2.2\%) is conservatively assigned to the full event fit 
by taking the largest deviation from the $\chi_{c2}$ decay mode 
with (without) a $\pi^0$ or $\eta$ decay.

The uncertainties associated with the detector resolution and 
minimum shower energy for the signal region are determined 
by varying these parameters in the study of $\chi_{c2}$ decays.  
The uncertainties are assigned by recalculating the $\chi_{c2} \to X$ 
efficiency-corrected yield  
and taking the difference from the nominal case.  
The detector resolution uncertainty is determined by individually varying 
the width of the core Gaussian and the transition point between the core Gaussian and 
the power law tail of the Crystal Ball function 
by 1 standard deviation, resulting in a total uncertainty of 1.0\%.  
An uncertainty of $3.2\%$ is assigned to the minimum photon energy requirement by 
increasing the lower bound from 90 to 110 MeV, corresponding to the same difference 
(18 MeV) between the minimum energy and the $\eta_c(2S)$ mass in the $\eta_c(2S)$ 
signal region.  

The uncertainties associated with selecting $\eta \to \pi^+ \pi^- \pi^0$ and 
$\eta^{\prime} \to \pi^+ \pi^- \eta$ decays based on the invariant mass of the decay 
products and the $\eta$ recoil mass used for $\psi(2S) \to \eta J/\psi$ 
suppression are determined by studying $\psi(2S) \to \gamma \chi_{c2}, \chi_{c2} \to X$ decays.  
The individual uncertainties are assessed by varying the 
respective mass range to double the detection inefficiency, 
redetermining the $\chi_{c2} \to X$ 
efficiency-corrected yield,  
and assigning the difference between this 
and the nominal result as the systematic uncertainty.  
An uncertainty of 1.3\% is assigned for the $\pi^+ \pi^- \pi^0$ invariant mass 
selection range in the $\pi^+ \pi^- \eta$ and $K^+ K^- \eta$ decay modes.  
An uncertainty of 1.2\% is assigned to the $\pi^+ \pi^- \eta$ invariant mass 
selection range for the  $\pi^+ \pi^- \eta^{\prime}$ decay mode.  
Uncertainties of 3.7\% and 0.2\% are assigned to the $\eta$ recoil mass 
suppression range for the $\eta \to \gamma \gamma$ 
and $\eta \to \pi^+ \pi^- \pi^0$ decays, respectively, 
comprising the $\pi^+ \pi^- \eta$ decay mode.

The largest systematic uncertainties in all $\eta_c(2S)$ decay modes 
arise from the $\Delta M$ selection criterion, 
the parameterization of the background shape, and 
the uncertainty in the $\eta_c(2S)$ mass, which are listed as separate entries in 
Tables \ref{table:NonEtaModeResults} and \ref{table:EtaModeResults}.  
They are estimated by varying the criterion being investigated, 
redetermining the 
efficiency-corrected yield  
upper limit in the $\eta_c(2S)$ signal region, 
and assigning the difference between this and the nominal result as the systematic 
uncertainty.  
The uncertainty associated with the $\Delta M$ criterion is determined by removing 
the cut.  
The uncertainty arising from the background modeling for the decay modes without 
an $\eta$ decay is determined by replacing the background determined from the 
background MC samples with a first-order polynomial.  The uncertainty arising from 
the background modeling for the decay modes with an $\eta$ decay is determined by 
lowering the overall background yield in the signal region by 1 standard deviation 
based on the data yield in the sideband region.
The effect of the uncertainty of the $\eta_c(2S)$ mass is determined by 
1 standard deviation variations of the mass, 
$M[\eta_c(2S)] = 3638 \pm 4~{\rm MeV}/c^2$, in the fits of the 
measured shower energy distributions and the determination of the detection 
efficiencies, with the larger discrepancy from the two cases being assigned as 
the systematic uncertainty.  

Additional systematic uncertainties in $\eta_c(2S)$ decay modes without an $\eta$ 
decay arise from the signal region range and the nonresonant component of the 
background, which are listed as separate entries in Table \ref{table:NonEtaModeResults}.  
The uncertainty associated with the maximum boundary of the signal region is 
assessed by varying the boundary by 8 MeV, with the larger deviation from nominal of 
the two cases being assigned as the systematic uncertainty.  The uncertainty 
associated with the minimum boundary of the signal region described above is 
combined in quadrature to obtain the signal region uncertainties listed in 
Table \ref{table:NonEtaModeResults}.
The uncertainty arising from the nonresonant component of the background 
was investigated by determining the ratio of event yields in the 20.6 pb$^{-1}$ 
of off-resonance data collected at $\sqrt{s}$ = 3.67 GeV with the yields 
from a 5 times luminosity continuum MC sample generated at the same center-of-mass 
energy.  The same event selection criteria were applied as for the $\eta_c(2S)$ 
signal search with the exception that $\Delta M$ was redefined as 
$\Delta M = 3.67~{\rm GeV} - M_{\rm inv}$.  
The uncertainty was assessed by repeating the fits of the $\eta_c(2S)$ signal 
region but with the continuum MC component of the background fixed 
to the ratio found in the off-resonance data sample study.  The difference 
between the $\eta_c(2S)$ signal yields from this and the nominal result was 
assigned as the systematic uncertainty.  No systematic uncertainties were applied 
to the $\eta$ decay modes due to the small amount of nonresonant background.

Tables \ref{table:NonEtaModeResults} and \ref{table:EtaModeResults} summarize 
the total systematic uncertainties.  
The individual uncertainties are treated as uncorrelated 
and are combined in quadrature to obtain the overall 
systematic uncertainties in the product branching fraction upper limits.  
The total uncertainty for composite decay modes is determined 
by weighting the total systematic uncertainty of each constituent decay mode 
by its branching fraction.

In addition to these sources of error, 
the partial width for a direct M1 radiative transition between $\psi(2S)$ and 
$\eta_c(2S)$ is related to the matrix element governing the spin-flip transition $I$ 
and the energy of the transition photon $E_{\gamma}$ by
\begin{equation}
\label{apptransitionrate}
\Gamma[\psi(2S)\to~\gamma\eta_c(2S)] \propto E_{\gamma}^3I^2.
\end{equation}  
This implies a signal shape given by a Breit-Wigner times $E_{\gamma}^3$ function, 
rather than the Breit-Wigner function that was used for our fits.  
We studied the effect of using this modified signal shape 
on the product branching fractions with the following procedure.  
For the determination of yields in $\eta_c(2S)$ modes without an $\eta$ decay, 
the signal regions were fitted with Breit-Wigner times $E_{\gamma}^3$ 
functions convoluted with Crystal Ball detector resolution signal shapes 
and the histogram backgrounds used in the nominal results.  
The nominal yields were used for modes with an $\eta$ decay.  
The efficiency for each mode was determined by applying the 
nominal event selection criteria to signal MC samples generated 
with signal shapes that were Breit-Wigner distributions multiplied by $E_{\gamma}^3$.  
Since there is no obvious choice of damping function as
there was for the ground state resonance \cite{Mitchell:CLEOjpsitoetac}, 
we use an arbitrary cutoff on the maximum allowed photon energy.  
With this procedure, we find deviations in the product branching fractions 
that are on the order of, and in some cases greater than, the other uncertainties.  
Because of the arbitrary cutoff, it is difficult to assign a systematic uncertainty 
to this effect, and we have chosen not to include it.

\section{Branching Fraction Results}

The upper limits on the number of signal events are used to set upper limits 
on the product branching fractions.  The product branching fraction 
for each $\eta_c(2S)$ decay mode is determined by Eq. (\ref{eq:lighthadronyields}), 
where $N_{sig}$ is the number of signal events and $N_{\psi(2S)} = 25.9~\times~10^6$.  
Because no statistically significant signals are observed 
in any of our ten decay channels, 
we use the efficiency-corrected yields 
to set upper limits on the product branching fractions.
Systematic uncertainties 
are determined and combined as described in Sec. \ref{sec:syst} 
and added to the statistically calculated product branching fraction upper limits.  
The final results are given in 
Tables \ref{table:NonEtaModeResults} and \ref{table:EtaModeResults}.

Upper limits for the product branching fraction are also determined as a function 
of the $\eta_c(2S)$ full width, for which the current world average is 
$\Gamma[\eta_c(2S)] = 14\pm7~{\rm MeV}/c^2$ \cite{PDG2008}.  
Separate signal MC samples with $\Gamma[\eta_c(2S)] =$ 7 and 21 MeV/$c^2$ 
were generated in the same manner as the nominal MC samples.  
The measured shower energy distributions are fitted in the same manner 
as for the standard yield determination procedure, 
but with the resolution functions determined from these MC samples 
and the full width of the signal shape adjusted to match the full width 
being investigated.  The linear extrapolation 
of the product branching fraction as a function of $\Gamma[\eta_c(2S)]$ 
is listed in Table \ref{table:WidthDepResults} for each $\eta_c(2S)$ decay mode. 

\begin{table}[htp]
\caption[WidthDepResults]
{\label{table:WidthDepResults}
Summary of product branching fraction results as a function of $\Gamma[\eta_c(2S)]$.  
The $y$-intercept and slope parameters $a$ and $b$ are defined by 
${\cal B}(\psi(2S) \to \gamma \eta_c(2S)){\cal B}(\eta_c(2S)\to X) 
< a + b\times\Gamma[\eta_c(2S)]$.  
Statistical and systematic uncertainties are included in these results.}
\begin{center}
\begin{tabular}{lcc}
  \hline
  \hline
          & $a$           & $b$                         \\
  Channel & $(10^{-6})$ & $(10^{-6}~c^2/{\rm MeV})$ \\
  \hline
  \hline 

  $K\bar{K}\pi$ & 6.6 & 0.56 \\ 

  $2(\pi^+\pi^-)$ & 6.5 & 0.58 \\ 

  $3(\pi^+\pi^-)$ & 4.0 & 0.74 \\ 

  $K^+K^-\pi^+\pi^-$ & 3.1 & 0.50  \\ 

  $K^+K^-\pi^+\pi^-\pi^0$ & 15.5  & 2.09 \\ 

  $K^+K^-2(\pi^+\pi^-)$ & 5.2 & 0.34 \\ 

  $K^0_SK^{\pm}\pi^{\mp}\pi^+\pi^-$ & 8.5 & 0.51 \\ 

  $\pi^+\pi^-\eta$ & 2.9 & 0.09  \\ 

  $K^+K^-\eta$ & 4.1 & 0.13 \\ 

  $\pi^+\pi^-\eta^{\prime}$ & 10.0  & 0.31 \\ 

  \hline
  \hline
\end{tabular}
\end{center}
\end{table}

\section{Search for \boldmath$\eta_c(2S) \to \pi^+ \pi^- \eta_c(1S)$}

\begin{figure}[htp]
\includegraphics*[width=3.8in]{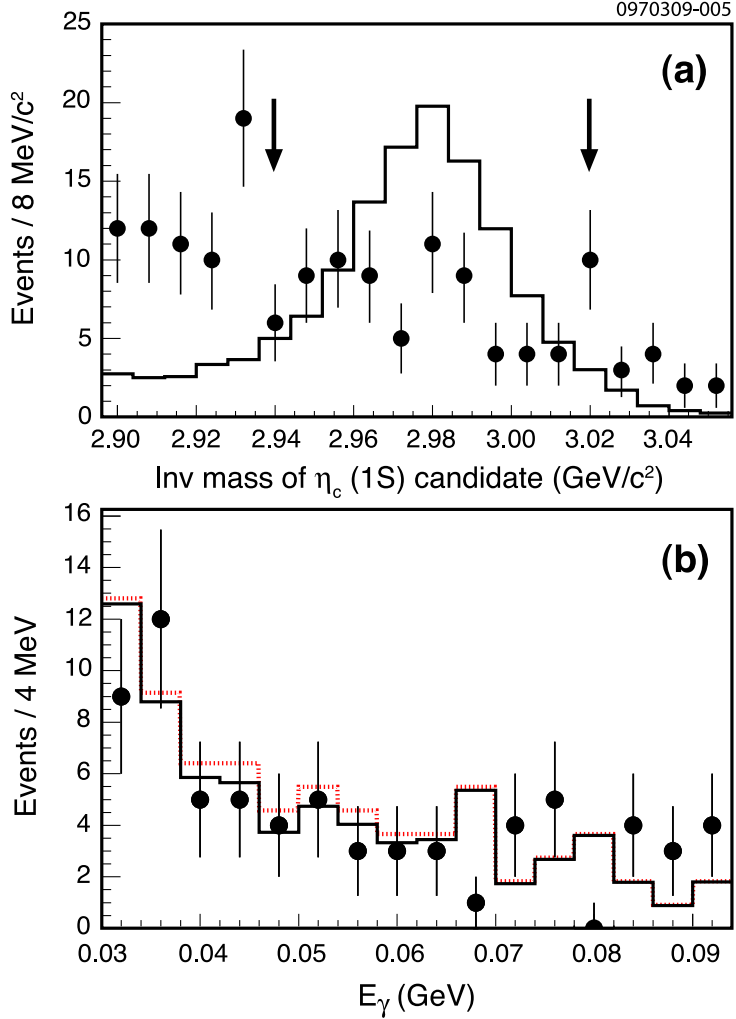}
\caption{(a) Hadronic invariant mass of $\eta_c(1S)$ candidates.  
The points are data; and the solid histogram is signal MC, 
arbitrarily normalized for clarity. 
Selected $\eta_c(1S)$ candidates are enclosed by the arrows.  
All other event selection criteria have been applied.  
(b) Measured shower energy distribution for candidates 
in the $\eta_c(2S)$ signal region 
after applying the selection criterion for the 
$\psi(2S) \to \gamma \eta_c(2S), \eta_c(2S) \to \pi^+ \pi^- \eta_c(1S)$ decay.  
The points are data; the dotted histogram is the background; 
and the solid histogram is the sum of signal and background.  
The signal is not shown since its area is less than zero.}
\label{fig:PiPiEtac}
\end{figure}

In addition to searching for $\eta_c(2S)$ decays to light hadrons, 
a search for the decay process 
$\psi(2S) \to \gamma\eta_c(2S), \eta_c(2S) \to \pi^+ \pi^- \eta_c(1S)$ is also 
performed.  
The four hadronic final states of $\pi^+\pi^-(K^0_S K^{\pm} \pi^{\mp})$, 
$\pi^+\pi^-(K^+ K^- \pi^0)$, $\pi^+\pi^-[2(\pi^+ \pi^-)]$, and 
$\pi^+ \pi^-(K^+ K^- \pi^+ \pi^-)$ 
plus a candidate transition photon 
are used for this study.  The selection criteria 
are the same as described above, except that the $J/\psi$ rejection criterion 
based on the $\pi^+ \pi^-$ recoil mass is removed and we require the hadronic 
decay products not associated with the dipion transition to be within 
40 MeV/$c^2$ of $M[\eta_c(1S)]$.  
Information from the $\pi^+ \pi^-$ recoil mass is not used since the distribution is 
broadened by the intrinsic widths of the $\eta_c(1S)$ and $\eta_c(2S)$.  
Figure \ref{fig:PiPiEtac}(a) shows the invariant mass of the $\eta_c(1S)$ candidates.  

Figure \ref{fig:PiPiEtac}(b) shows the measured shower energy 
distribution after applying the $\eta_c(2S) \to \pi^+ \pi^- \eta_c(1S)$ 
selection criterion.  No evidence of a signal is observed.  
The $\chi^2$/d.o.f. of the fit is 19.5/14.  The number of 
signal events is $N_{sig} = -5.1^{+10.7}_{-9.8}$, corresponding to an 
upper limit of $N_{sig} < 14.8$ (90\% C.L.). 
   
The upper limit on the product branching fraction is determined by 
\begin{equation}
{\cal B}(\psi(2S) \to \gamma \eta_c(2S)){\cal B}(\eta_c(2S)\to \pi^+\pi^-\eta_c(1S))
= \frac{N_{sig}}{N_{\psi(2S)}[\sum_i \epsilon_i \times {\cal B}_i(\eta_c(1S))]},
\label{eq:br4pipietac1s}
\end{equation}
where, for a given final state $i$, $\epsilon_i$ is the detection efficiency 
and ${\cal B}_i(\eta_c(1S))$ is the branching fraction for the $\eta_c(1S)$ decay.  
The signal efficiency for each $\eta_c(1S)$ decay mode is determined from 
signal MC samples. The decay $\psi(2S)\to \gamma\eta_c(2S)$ 
is generated in the same manner as described in Sec. \ref{sect:datasamANDevtsel}.  
The decay $\eta_c(2S) \to \pi^+ \pi^- \eta_c(1S)$ is generated according to phase 
space with the $\eta_c(1S)$ parameters $M[\eta_c(1S)] = 2979.8 \pm 1.2~{\rm MeV}/c^2$ 
and $\Gamma[\eta_c(1S)] = 26.5 \pm 3.5~{\rm MeV}/c^2$ \cite{PDG2007}.  
The individual values used to determine 
$\sum_i \epsilon_i \times {\cal B}_i(\eta_c(1S)) = (0.50\pm0.07)\%$ 
are listed in Table \ref{table:pipieceff}.

\begin{table}[htbp]
\caption{\label{table:pipieceff}
Efficiency and submode branching fraction information for the 
$\psi(2S) \to \gamma \eta_c(2S), \eta_c(2S) \to \pi^+ \pi^- \eta_c(1S)$ study.  
We assume $\Gamma[\eta_c(2S)] = 14~{\rm MeV}/c^2$ \cite{PDG2008,PDG2007}.  The column 
${\cal B}(\eta_c(1S))$ lists the branching fractions for the $\eta_c(1S)$ decay \cite{PDG2008,PDG2007}.  
The listed efficiencies include submode branching fractions.}  
\begin{center}
\begin{tabular}{lccc}
  \hline
  \hline
  $\eta_c(1S)$ Decay Mode & ~~~$\epsilon$ (\%)~~~ & ~~~${\cal B}(\eta_c(1S))$ (\%) 
                          & ~~~$\epsilon\times{\cal B}(\eta_c(1S))$ (\%) \\
  \hline
  \hline
  $K^0_SK^{\pm}\pi^{\mp}$ &  $6.64\pm0.14$ & $2.3\pm0.4$   & $0.148\pm0.026$ \\
  $K^+ K^- \pi^0$         &  $7.33\pm0.18$ & $1.17\pm0.20$ & $0.086\pm0.015$ \\
  $2(\pi^+\pi^-)$         & $10.99\pm0.21$ & $1.2\pm0.3$   & $0.13\pm0.03$ \\
  $K^+K^-\pi^+\pi^-$      &  $8.88\pm0.20$ & $1.5\pm0.6$   & $0.13\pm0.05$ \\
  \hline
  \hline
\end{tabular}
\end{center}
\end{table}

Several sources of systematic uncertainty in the 
$\eta_c(2S) \to \pi^+ \pi^- \eta_c(1S)$ measurement have been evaluated.  
The uncertainty due to the invariant mass range used to select $\eta_c(1S)$ 
candidates is determined by tightening the mass range to double the detection 
inefficiency, redetermining the product branching fraction upper limit, and 
assigning the relative difference between this and the nominal result 
as the systematic uncertainty.  
The effect of the uncertainty of the $\eta_c(1S)$ full width is determined by 
generating separate signal MC samples with $\Gamma[\eta_c(1S)] =$ 23 and 30 MeV, 
{\it i.e.}, 1 standard deviation variations of 
$\Gamma[\eta_c(1S)] = 26.5 \pm 3.5~{\rm MeV}/c^2$ \cite{PDG2007}, 
to redetermine the detection efficiencies and repeating the yield determination 
procedure with the resolution functions determined from these MC samples.  
The uncertainties associated with the $\Delta M$ criterion, background 
parameterization, $M[\eta_c(2S)]$ uncertainty, signal region, and nonresonant 
background component are evaluated by performing the procedures described in 
Sec. \ref{sec:syst}.  
The systematic uncertainties associated with the number of $\psi(2S)$ decays, 
trigger efficiency, particle reconstruction, PID, full event and vertex fitting, 
and the suppression of transitions to other charmonium states 
for these specific final states are assigned as described in Sec. \ref{sec:syst}.  
Table \ref{table:pipiecsyst} lists the 
individual contributions to the total systematic uncertainty of the product 
branching fraction.

\begin{table}[htbp]
\caption{\label{table:pipiecsyst}
Sources of systematic uncertainties for the product branching fraction 
${\cal B}(\psi(2S) \to \gamma~\eta_c(2S)){\cal B}(\eta_c(2S) \to \pi^+ \pi^- \eta_c(1S))$.  
The total systematic uncertainty is determined by combining the individual 
contributions in quadrature.}
\begin{center}
\begin{tabular}{lc}
  \hline
  \hline
  Source & Uncertainty (\%) \\
  \hline
  \hline
  Nonresonant background & 28.6 \\
  $M[\eta_c(2S)]$ & 22.0 \\ 
  Signal region & 20.7 \\
  ${\cal B}(\eta_c(1S))$ & 14.0 \\
  $M[\eta_{c}(1S)]$ & 11.1 \\
  $\Delta M$ & 7.9 \\
  $\Gamma[\eta_c(1S)]$ & 4.1 \\
  Background parameterization & 1.5 \\
  Other & 5.5 \\ \hline
  Total & 46.5 \\ 
  \hline
  \hline
\end{tabular}
\end{center}
\end{table}

The product branching fraction for 
$\psi(2S) \to \gamma \eta_c(2S),\eta_c(2S) \to \pi^+ \pi^- \eta_c(1S)$, 
assuming $\Gamma[\eta_c(2S)] =  14~{\rm MeV}/c^2$ and 
including statistical and systematic uncertainties, is 
${\cal B}(\psi(2S) \to \gamma \eta_c(2S)){\cal B}(\eta_c(2S) \to \pi^+ \pi^- \eta_c(1S)) 
= (-0.39^{+0.83}_{-0.76}\pm0.18)\times 10^{-4} 
< 1.7 \times 10^{-4}$ (90\% C.L.).  
Expressed as a function of $\Gamma[\eta_c(2S)]$, the upper limit is 
$\{(48) + (9.2~c^2/{\rm MeV})\times\Gamma[\eta_c(2S)]\}\times 10^{-6}$.

\section{Summary and Conclusions}

In summary, we do not observe the transition $\psi(2S) \to \gamma \eta_{c}(2S)$ 
with any of the ten exclusive $\eta_c(2S)$ decays to light hadrons.  
We also do not observe evidence for the decay process 
$\psi(2S) \to \gamma \eta_c(2S), \eta_c(2S) \to \pi^+ \pi^- \eta_c(1S)$. 
Our original objectives for measuring the properties of the $\eta_c(2S)$ 
cannot be achieved with this data sample, and only upper limits 
for the product branching fractions are obtained.

The BaBar Collaboration recently reported a branching fraction of 
${\cal B}(\eta_{c}(2S) \to K{\bar K}\pi) 
= (1.9\pm0.4(stat)\pm1.1(syst))\%$ \cite{Aubert:2008kp}, 
where the systematic uncertainty is dominated by the inclusive measurement of 
$B^{\pm} \to K^{\pm} \eta_{c}(2S)$ \cite{InclEtac2S}.  
Using the central value of ${\cal B}(\eta_{c}(2S) \to K{\bar K}\pi)$ 
and our 90\% confidence level upper limit of
$\mathcal{B}(\psi(2S) \rightarrow \gamma \eta_c(2S))
\mathcal{B}(\eta_c(2S)\rightarrow K \bar{K} \pi) < 14.5\times10^{-6}$ 
leads to $\mathcal{B}(\psi(2S)\rightarrow \gamma \eta_c(2S)) < 7.6 \times 10^{-4}$, 
which is larger than the phenomenological prediction of 
${\cal B}(\psi(2S) \to \gamma \eta_c(2S)) = (3.9 \pm 1.1)\times 10^{-4}$.

The $\mathcal{B}(\eta_c(2S) \rightarrow K \bar{K} \pi)$ measurement 
can also be used to determine upper limits of $\eta_c(2S)$ hadronic decays 
based on published $\eta_c(2S)$ searches.  
The two-photon fusion result reported by the CLEO Collaboration \cite{cleo2004fusion} 
and the $\mathcal{B}(\eta_c(2S) \rightarrow K \bar{K} \pi)$ measurement 
lead to a two-photon partial width of $\Gamma_{\gamma \gamma}[\eta_c(2S)]= 4.8 \pm 3.7$~keV.  
Using this value of $\Gamma_{\gamma \gamma}[\eta_c(2S)]$ 
with the recent two-photon fusion upper limits for $\eta_c(2S)$ production 
from the Belle Collaboration \cite{BelleGG}, we find 
$\mathcal{B}(\eta_{c}(2S) \to 2(\pi^+\pi^-)) < 0.14\%$ 
and $\mathcal{B}(\eta_{c}(2S) \to K^+K^-\pi^+\pi^-) < 0.10\%$ (90\% C.L.).  
These upper limits are an order of magnitude smaller than the branching fractions 
obtained by assuming that the partial widths 
for $\eta_c(2S)$ decays are the same as for $\eta_c(1S)$, 
{\it i.e.}, $\mathcal{B}(\eta_{c}(2S) \to 2(\pi^+\pi^-)) = (2.3\pm0.6\pm1.2)\%$ and 
$\mathcal{B}(\eta_{c}(2S) \to K^+K^-\pi^+\pi^-) = (2.9\pm1.1\pm1.5)\%$, 
where the first error is the uncertainty from the $\eta_c(1S)$ branching fraction 
and the second error is the uncertainty from $\Gamma[\eta_c(2S)]$ \cite{PDG2008}.

\section{Acknowledgments}

We gratefully acknowledge the effort of the CESR staff
in providing us with excellent luminosity and running conditions.
D.~Cronin-Hennessy and A.~Ryd thank the A.P.~Sloan Foundation.
This work was supported by the National Science Foundation,
the U.S. Department of Energy,
the Natural Sciences and Engineering Research Council of Canada, and
the U.K. Science and Technology Facilities Council.


\end{document}